# Experimental features of emissions and fuel consumption in a car-following platoon

Shirui Zhou[1,2], Ying-En Ge[3], Shaowei Yu[3], Junfang Tian[1,2*], Rui Jiang[4]*

[1]Institute of Systems Engineering, College of Management and Economics, Tianjin University, No. 92 Weijin Road, Nankai District, Tianjin 300072, China

[2]Laboratory of Computation and Analytics of Complex Management Systems (CACMS), Tianjin University, Tianjin, 300072, China

[3]College of Transportation Engineering, Chang'an University, Xi'an, Shaanxi, 710064, China

[4]Key Laboratory of Transport Industry of Big Data Application Technologies for Comprehensive Transport, Ministry of Transport, Beijing Jiaotong University, Beijing 100044, China

* Corresponding authors.

E-mail addresses: jftian@tju.edu.cn (J. Tian), jiangrui@bjtu.edu.cn (R. Jiang).

# Experimental features of emissions and fuel consumption in a car-following platoon

## Abstract

The paper investigates the features of emissions and fuel consumption (EFC) in a car-following (CF) platoon based on two experimental datasets. Four classical EFC models are employed and a universal concave growth pattern of the EFC along a platoon has been demonstrated. A general framework of coupling EFC and CF models is tested by calibrating and simulating three classical CF models. This work first demonstrates that, at vehicle-pair level, all models perform well on EFC prediction. The intelligent driver model outperforms the other CF models on calibration accuracy, but this is not true on EFC prediction. Second, at platoon level, the predicted EFC is nearly constant along the platoon which qualitatively differs from the experimental observation. The investigation highlights that accurate estimations at vehicle level may be insufficient for analysis at platoon level due to the significant role of oscillation growth and evolution in EFC estimation.

# 1 Introduction

Accurately assessing individual emissions and fuel consumption (EFC) and capturing the property of EFC in complex transportation environments is a prerequisite for evaluating environmental impact and conducting sustainable traffic management. For decades, the predominant approach for achieving this has been the coupling of microscopic traffic flow models (e.g., Bando et al., 1995; Gipps and Wilson, 1980; Jiang et al., 2001; Newell, 2002; Treiber et al., 2000) with EFC models (e.g., Ahn et al., 2002; Ahn and Rakha, 2009; Barth and Boriboonsomsin, 2009; Hajmohammadi et al., 2019; Lei et al., 2010).

Both types of models are designed to capture the non-linearity of input and output. Microscopic traffic flow models, primarily car-following (CF) models, aim to comprehensively and accurately depict driving behavior in terms of physics and psychology, and thus produce outputs such as velocity and acceleration. The recent availability of abundant trajectory data has significantly contributed to the development of microscopic CF models, which forms the foundation of the EFC prediction framework, see Figure 1 below.

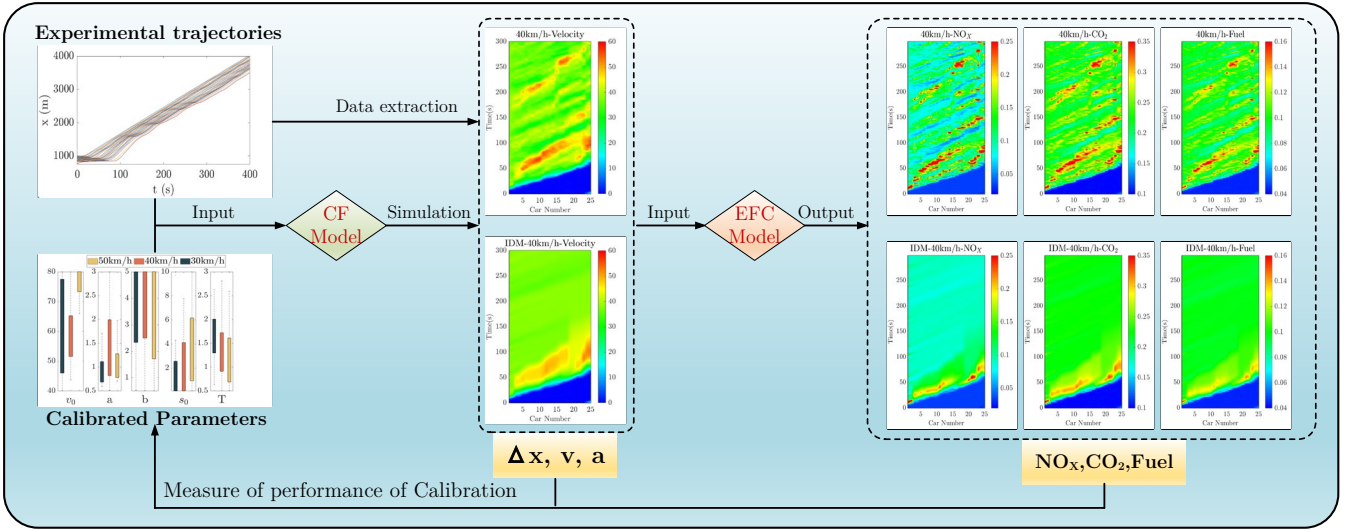

Figure 1: A general coupling framework of EFC calculation and prediction.

All elements in Figure 1 are sampled from this paper to illustrate the structure of the general coupling framework of EFC calculation and prediction. The propagation of errors from an input and output process increases the complexity of estimating EFC accurately. The CF and EFC models serve the core components of this framework and require development, particularly for platoon EFC prediction. Furthermore, the selection of performance measures for calibration, whether single-objective or multi-objective, CF-related or EFC-related, necessitates careful consideration, as discussed by Punzo et al. (2021). These topics are out of the scope of this paper, and will be investigated in future work.

However, collecting EFC data is more challenging and costly than trajectory data (Ma et al., 2014). To overcome these limitations, recent efforts have focused on conducting broad measurements of vehicle EFC to develop aggregating average EFC models (Ahn et al., 2002;

Akcelik, 1989) that can avoid the complex calibration procedures required by EFC models. This has established the coupling EFC prediction framework. By incorporating various specific EFC factors (see Subsection 2.2), EFC models can map vehicle states (e.g., real-time velocity and acceleration) to EFC. Therefore, precise EFC calculations can only be achieved if the output of a real-world or reliable traffic flow model is input into the EFC models (Song et al., 2013).

Before employing the coupled framework of EFC models and CF models, the calibration procedure of the CF models is deemed indispensable. The calibration of traffic flow models is a complex and well-developed area. Notable advances have been made to enrich the methodology (Ciuffo et al., 2014; Hoogendoorn and Hoogendoorn, 2010; Punzo et al., 2021; Punzo and Montanino, 2016; Sharma et al., 2019; Treiber and Kesting, 2013). In the context of EFC estimation, the calibration of traffic flow models against individual trajectories and the combination of them with the aggregation of average EFC models to estimate or predict EFC has been extensively applied. Several practices for the prediction of EFCs based on the calibration of the coupled framework are available. The most straightforward approach is to directly apply the built-in models and calibration modules in the simulation software. For instance, Chauhan et al. (2019) proposed an EFC calculation framework for signalizing intersections by directly calibrating and calculating EFC with the integrated modules in VISSIM.

In addition, various approaches have been used to calibrate CF models to estimate the EFC. One common approach is to estimate optimal parameters by minimizing the errors in spacing or velocity over the entire trajectory. Subsequently, the EFC can be estimated by using the EFC model with the output of the calibrated CF models. For example, Vieira da Rocha et al. (2015) studied the impact of traffic-related factors on the calibration and the EFC estimation by calibrating the single trajectory on velocity errors. They found that non-optimal parameters significantly increase the uncertainty of EFC output in a single trajectory but have less impact on platoons. Meng et al. (2021) calibrated a new multidimensional stochastic NCM for EFC predictions using the DTW algorithm to identify the similarity between the simulated and observed speed time series. Jie et al. (2013) sought optimal parameters by minimizing the difference between the observed and simulated speed and acceleration distributions to obtain accurate input for EFC models. Another approach is to minimize errors on EFC-related indexes, such as the VSP distribution. Wang et al. (2017) calibrated three CF models against empirical data to evaluate their ability to reproduce the VSP distributions. This method was also used by Song et al. (2013) to analyze the VSP distributions generated by the calibration results of different CF models. They found that the calibration might not improve the EFC errors by global sensitivity analysis. In addition to the traditional calibration objective function, Song et al. (2015) adopted the differences between the simulated and empirical VSP distributions as the calibration objective function and successfully generated realistic VSP distributions.

Although the two types of models can be developed relatively independently and conveniently coupled, the errors and uncertainty may be further heightened in the coupling process, which raises the complexity of modeling and calibration (Zhou et al., 2016). However, most of the above practices of EFC predictions are based on the vehicle level and adopt the trajectory-related or EFC-related index as the performance measure. Hence, these works mostly relate to the local predictions that focus on the vehicle pair.

In recent years, owing to vehicle-to-vehicle (V2V) development and other emerging technology, connected autonomous vehicles are being deployed and used to improve traffic stability and EFC under congested platoons. Wang et al. (2015) proposed an instantaneous control model to suppress the traffic oscillation and significantly reduced the EFC in platoons in stop-to-go traffic situations. Zhang et al. (2018) analyzed the impact of safety assistance driving systems (SADS) on EFC in a car platoon. Results show that incorporating a two-regime SADS with different decelerations in each regime can increase traffic stability and decrease the EFC. All these advanced researches highlight the deep connection between the traffic instability of platoons and EFC predictions.

Nevertheless, as Punzo and Montanino (2020) pointed out, the calibration results against individual trajectories may not be able to reproduce systematic patterns. In real traffic, the oscillations usually propagate and evolve along the car platoons, which causes considerable additional EFC. Recent efforts were reported (see, e.g., Huang et al., 2018; Jiang et al., 2018, 2014; Tian et al., 2021, 2019; Zheng et al., 2022, 2021) in which CF experiments are conducted and the traffic instability in the platoon is revealed. Therefore, the investigation on EFC should be performed not only on the vehicle-pair level but also on the platoon level.

However, to our knowledge, Vieira da Rocha et al. (2015) conducted the only platoon-level calibration study on EFC. They investigated the EFC of cars in the 10-car platoons on a 500 m length section extracted from NGSIM data. They found that (i) the optimal parameters of the CF rules are insufficient to get accurate estimates of the EFC at a vehicle level; (ii) However, optimal or even averaged traffic parameters seem sufficient to estimate the EFC at a platoon level. For the 10-car platoons, the trajectory of the last car is similar to that of the leading one, indicating the traffic oscillations are nearly fully developed. Moreover, the duration of the data is only less than 120 s. As a result, the oscillation growth and evolution are not well captured, and the role of oscillation growth and evolution in estimating EFC remains unclear. This study is motivated by the research gap. Based on our recent experimental data, we study the EFC in a 25-car platoon, in which the leading car moves nearly constantly. As a result, the formation and growth of traffic oscillations can be observed. In addition, the EFC in a 40-car platoon on a circular road is also investigated where the platoon size is larger and the traffic oscillations are fully developed.

We aim to investigate the following questions:

i. How does the EFC evolve along the car platoon with the growth of traffic oscillations?

ii. Are the optimal parameters of the CF rules sufficient to accurately estimate the EFC of a long platoon at the vehicle level and the platoon level?

To this end, four classical EFC models are applied for the analysis of the empirical feature of EFC along the car platoon. Three typical microscopic traffic models are coupled with the EFC model and calibrated. Two simulation tests are conducted from the local and platoon levels using calibrated models, respectively.

This paper is structured as follows: Section 2 provides a brief overview of CF models and EFC models. Section 3 presents the empirical property of EFC along the platoon based on our data and four different EFC models. Section 4 focuses on the calibration and simulations of three CF models, and the results in terms of EFC prediction are discussed. Finally, Section 5 concludes the paper.

## 2 Models

This section introduces the three classical CF models and the state-of-the-art EFC models.

### 2.1 Car-following models

For convenience to be coupled with EFC models, microscopic car-following models with moderate parameters and reliable representations of driving behavior are used. Many contributions have been made to generate realistic acceleration and develop classical CF models (e.g., Bando et al., 1995; Gipps and Wilson, 1980; Jiang et al., 2001; Newell, 2002; Treiber et al., 2000). Additionally, the Wiedemann model and Fritzsche model are other commonly used models which have been deeply investigated by Song et al. (2015) and are based on the division of CF regimes. This paper selects three classical CF models to calibrate and validate the platoon EFC, i.e., the Full Velocity Difference model (FVDM), the Intelligent Driver model, and Gipps' model (Gipps). So the three models will be briefly reviewed in this section.

### 2.1.1 Intelligent driver model (IDM)

As a far-famed classical traffic flow model, the parameters in the intelligent driver model (IDM) have intuitive physical meanings and each corresponds to one driving regime (Treiber et al. 2000). Massive modification versions of IDM have been developed for different research needs. IDM has also been embedded in many simulation software (e.g. SUMO, VISSIM). Considering the wide use and simple structure of the IDM model, we choose it as one typical CF model.

In the IDM, the acceleration of a vehicle (numbered $n$) is determined by:

$$a_n(t) = a_{\max}\left(1 - \left(\frac{v_n(t)}{v_{\max}}\right)^4 - \left(\frac{s_0 + v_n(t)T - \frac{v_n(t)\Delta v_n(t)}{2\sqrt{a_{\max}b}}}{\Delta x_n(t)}\right)^2\right) \quad (1)$$

where $a_n(t)$ and $v_n(t)$ are respectively the acceleration and velocity of the $n^{th}$ vehicle at time $t$, $a_{\max}$ is the maximum acceleration, $b$ is the comfortable deceleration, $v_{max}$ is the maximum speed, $s_0$ is the bumper-to-bumper gap in jams, $T$ is the desired time gap in each simulation step, $\Delta v_n(t) = v_{n-1}(t) - v_n(t)$ is the velocity difference, $\Delta x_n = x_{n-1}(t) - x_n(t) - l_{\text{veh}}$ is the distance between the $n^{th}$ vehicle and its leading vehicle $(n-1)^{th}$ and $l_{\text{veh}}$ is the length of the vehicle.

### 2.1.2 Full velocity difference model (FVDM)

The full velocity difference model (FVDM) was proposed by Jiang et al. (2001)) and belongs to the optimal velocity model family, which considers the impact of velocity difference in CF. The acceleration of vehicles is determined by the following:

$$a_n(t) = \kappa[V(\Delta x_n(t)) - v_n(t)] + \lambda(\Delta v_n(t)) \quad (2)$$

$$V(\Delta x_n(t)) = \frac{v_{\max}}{2}(\tanh(\zeta \Delta x_n(t) - \beta) + \tanh(\beta)) \qquad (3)$$

where $\kappa, \lambda, \zeta$ and $\beta$ are sensitivity coefficients, $V(\Delta x_n(t))$ is the optimal velocity and $v_{\max}$ is the maximum velocity.

### 2.1.3 Gipps model (Gipps)

Gipps' car-following model (Gipps and Wilson 1980) focuses on the driving safety condition by incorporating reaction time. The velocity is derived from the vehicular kinematic equation to ensure collision avoidance under the current reaction delay. The formulation of Gipps' for this purpose is as follows:

$$v_n(t+\tau) = \min(v_n(t) + a_{max}\tau, v_{\text{safe}}) \qquad (4)$$

$$v_{\text{safe}} = -b\tau + \sqrt{b^2\tau^2 + 2b(x_{n-1}(t-\tau) - x_n(t-\tau) - s_0) + v_{n-1}^2(t-\tau)} \qquad (5)$$

where $\tau$ is the time delay and spacing delay, respectively, and the difference in velocity calculates acceleration.

## 2.2 An overview of EFC models

According to a recent review (Zhou et al., 2016), six core factors impact EFC: travel-related, weather-related, vehicle-related, roadway-related, and driver-related factors. On this basis, various models are proposed to estimate the EFC (e.g., Ahn et al., 2002; Ahn and Rakha, 2009; Barth and Boriboonsomsin, 2009; Hajmohammadi et al., 2019; Lei et al., 2010) while classification criteria are multi-dimensional. Regarding transparency, they can be divided into white-box, gray-box, and black-box models. Meanwhile, they can be distinguished further by model inputs as engine-based, vehicle-based, and modal-based models. In this paper, four models are selected, that is, the VT-micro model, MEF model, VSP model, and ARRB model, for our analysis given that:

(i) The four models are all black-box models regarding a simple formulation and few parameters and only require information on acceleration and velocity.

(ii) They are in diversity to ensure the universality of the following analysis results. The VT-micro and MEF models are vehicle-based. The VSP and ARRB models are modal-based.

(iii) They all focus on the two core factors related to traffic oscillations, that is, vehicle-related and driver-related factors. Hence, they may be suitable for reflecting the EFC of CF dynamics.

### 2.2.1 VT-Micro model

VT-Micro model (Ahn et al., 2002) has reasonable estimation accuracy and was validated with field data. It is given with the exponential function as the measure of effectiveness (MOE),

$$MOE(a_n(t), v_n(t)) = e^{P(a_n(t), v_n(t))} \tag{6}$$

where the exponent $P$ is a polynomial function of speed and acceleration,

$$P(a_n(t), v_n(t)) = \sum_{i=0}^{3} \sum_{j=0}^{3} K_{ij} (v_n(t))^i (a_n(t))^j \tag{7}$$

$K_{ij}$ are the regression coefficients from field measurements. Ahn et al. (2002) have calibrated the corresponding regression coefficients $K_{ij}$ of the fuel consumption, the $CO_2$ emission, and the $NO_x$ emission by using the data collected at the Oak Ridge National Laboratroy (see Appendix A). In addition to its precise prediction performance, the VT-micro model also effectively distinguishes different operation modes by adopting different coefficients for positive and negative acceleration.

### 2.2.2 MEF model

As an extended version of the VT-micro model, the MEF model incorporates historical acceleration information into the calculation formulation (Lei et al., 2010). By linearly combining the current and historical acceleration, the fuel consumption of several types of trucks under high-speed conditions can be precisely estimated. The composited acceleration is as follows:

$$\bar{a}_n(t) = \alpha \cdot a_n(t) + (1-\alpha)\sum_{i=1}^{9} a_n(t-i)/9 \tag{8}$$

so that the polynomial function is:

$$P(\bar{a}_n(t), v_n(t)) = \sum_{i=0}^{3}\sum_{j=0}^{3} K_{ij}(v_n(t))^i \left(\alpha \cdot a_n(t) + (1-\alpha)\sum_{i=1}^{9} a_n(t-i)/9\right)^j \tag{9}$$

The value of parameters is the same as the VT-micro model.

### 2.2.3 ARRB model

ARRB model is one of the classic modal-based models (Akcelik, 1989). The fuel consumption is calculated by considering all operation modes, including constant-speed driving, accelerating, etc. It could be expressed as:

$$f = \alpha + \beta_1(d_1 v + d_2 v^3 + d_3 v^2 + mav) + \beta_2 m\max(0,a)^2 v \tag{10}$$

where the parameters set for the Cortina test car are $\begin{cases} \alpha = 0.666 mL/s, m = 1680 kg \\ \beta_1 = 0.072, \beta_2 = 0.033984 \\ d_1 = 0.269, d_2 = 0.000672, d_3 = 0.0171. \end{cases}$

### 2.2.4 VSP model

The VSP model is another modal-based model (Duarte et al., 2015; Jimenez-Palacios, 1998). It introduces the intermediate variable Vehicle Specific Power based on the forces on one vehicle into the fuel consumption estimation process.

$$VSP_i = v \cdot (1.1 \cdot a + 9.81 \cdot grade + 0.132) + 3.02 \cdot 10^{-4} \cdot v^3 \quad (11)$$

where the grade is road slope and set to zero in this paper. Regarding the specific relationship between fuel consumption and VSP, there are three zones of the corresponding equations:

$$F = \begin{cases} f, VSP_i < -10 \\ aVSP_i^2 + bVSP_i + c, 10 \leq VSP_i \leq -10 \\ mVSP_i + d, VSP_i > -10 \end{cases} \quad (12)$$

And the values of the parameters are: $\begin{cases} f = 2.48 \cdot 10^{-3} \\ a = 1.98 \cdot 10^{-3}, b = 3.97 \cdot 10^{-2}, c = 2.01 \cdot 10^{-1} \\ m = 7.93 \cdot 10^{-2}, d = 2.48 \cdot 10^{-3}. \end{cases}$

# 3 Experimental properties of EFC

## 3.1 Experiment I: 25-car platoon on a straight road

### 3.1.1 Experimental setup

The 25-car-platoon experiment was performed on a 3.2 km long non-signalized stretch in Hefei City, China (Jiang et al., 2015, 2014). We have equipped all the cars with high-precision differential GPS devices. The velocities and locations were recorded every 0.1 s. In the experiment, we asked the leading car driver to drive at several pre-given constant speeds. Other drivers were asked to follow each other and drive their cars as usual. Overtaking is forbidden. The car platoon decelerates, makes a U-turn, and stops when reaching the road section's end. When all the cars have stopped, a new run of the experiment begins. We have carried out two sets of experiments in which the sequence of the cars (mainly the velocity of the leading vehicle) has been changed. There are 12 runs in Set 1 and 23 runs in Set 2. Details of the runs in each set of the experiment are also presented. For more details, one can refer to Jiang et al. (2014). This experiment reproduces the CF dynamics in the scenery of a moving bottleneck on a straight road with velocity oscillations. Given in the U-turn and deceleration process it is no longer CF, they are deleted in the trajectories in the following analysis.

### 3.1.2 Spatiotemporal properties in the 25-car platoon

This experiment performs one of the most common scenarios of CF platoons on the freeways. This paper selects three typical runs with different $v_{\text{leading}}$ presenting different magnitudes of velocity perturbations and spatiotemporal characteristics for analysis (see Figure 2). The velocity of the leading vehicle oscillates around the pre-given values with a small perpetuation because of the driving behavior stochasticity. Moreover, the following velocity oscillation is growing and propagating along the platoon. With different $v_{\text{leading}}$, the platoon shows quantitatively different spatiotemporal properties on velocity. Hence, the EFC properties under three $v_{\text{leading}}$ may be different.

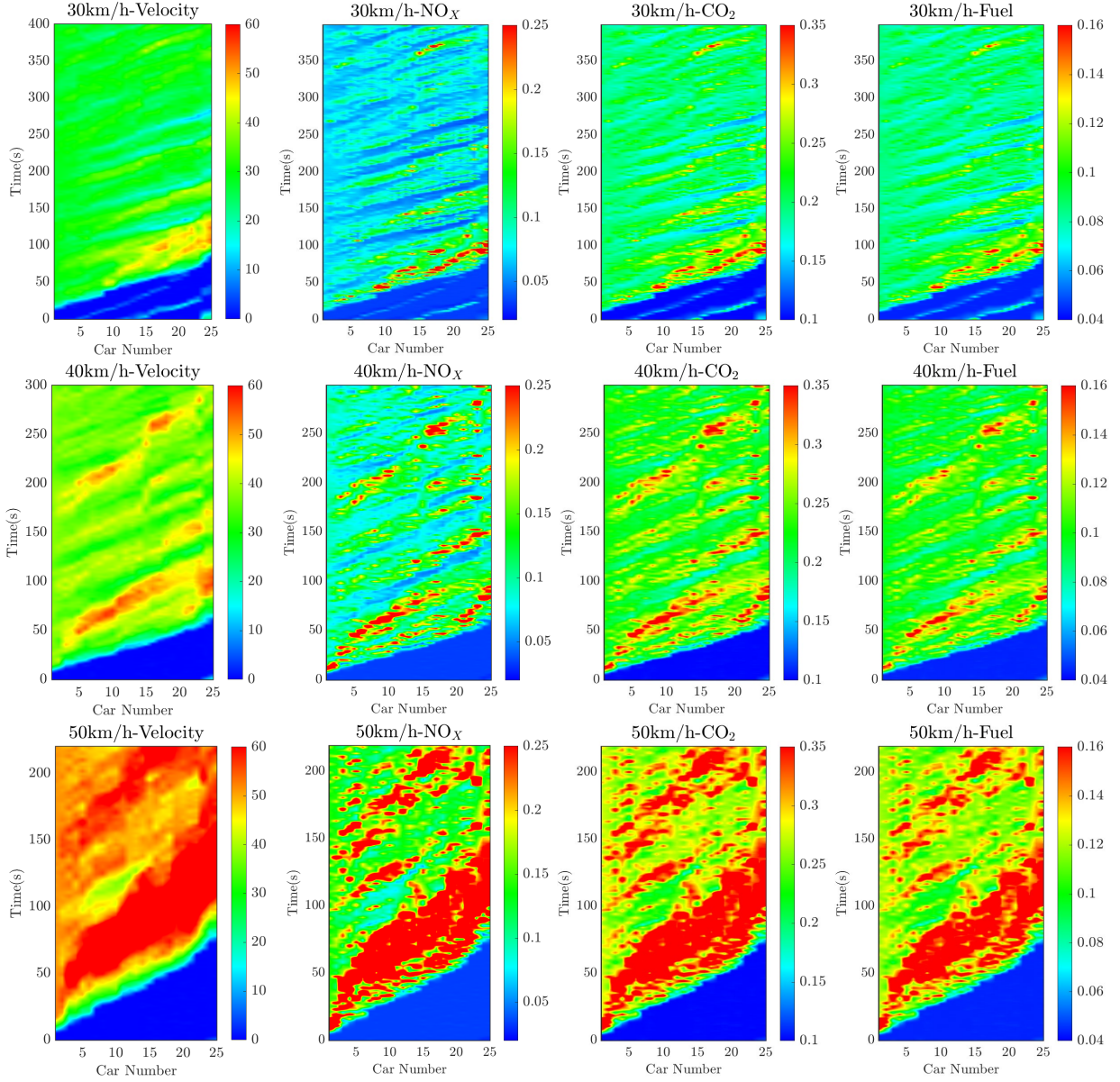

Figure 2: Evolution of the spatiotemporal pattern of velocity and EFC in the experiment. EFC is calculated by the MEF model. From left to right, these panels represent velocity (km/h), $NO_x$ (mg), $CO_2$ (g), and Fuel (ml), respectively.

By using four EFC models, we calculated the instant EFC of the platoons. Figure 2 shows the calculated $NO_X$, $CO_2$, and Fuel by MEF model. It can demonstrate that the spatiotemporal pattern of EFC is greatly correlated to velocity. Moreover, the profile of spatiotemporal patterns of three EFC indices $NO_X$ $CO_2$ and $Fuel$ are very similar. It indicates the inner consistency of the EFC model in estimating different EFC indices so that the spatiotemporal patterns can also be preserved.

### 3.1.3 Growth pattern along the 25-car platoon

According to Jiang et al. (2015, 2014) and Tian et al. (2019, 2016), the standard deviations of velocity grow in a concave way in the CF platoons which is a universal mechanism of traffic oscillations. It means the oscillation may increase with the increase in car number, which is proven to the universal in terms of traffic flow instability. From Subsection 2.2, It can be found that most EFC models rely on accurate velocity and acceleration so that the growth of traffic oscillations will have a great impact on EFC. As for our experimental data, the propagation of speed oscillations may lead to different EFCs along the car platoons. Hence, to accurately assess the EFC properties along the platoon and estimate it, the empirical property of the EFC of each vehicle should be considered as they are directly related to speed and acceleration.

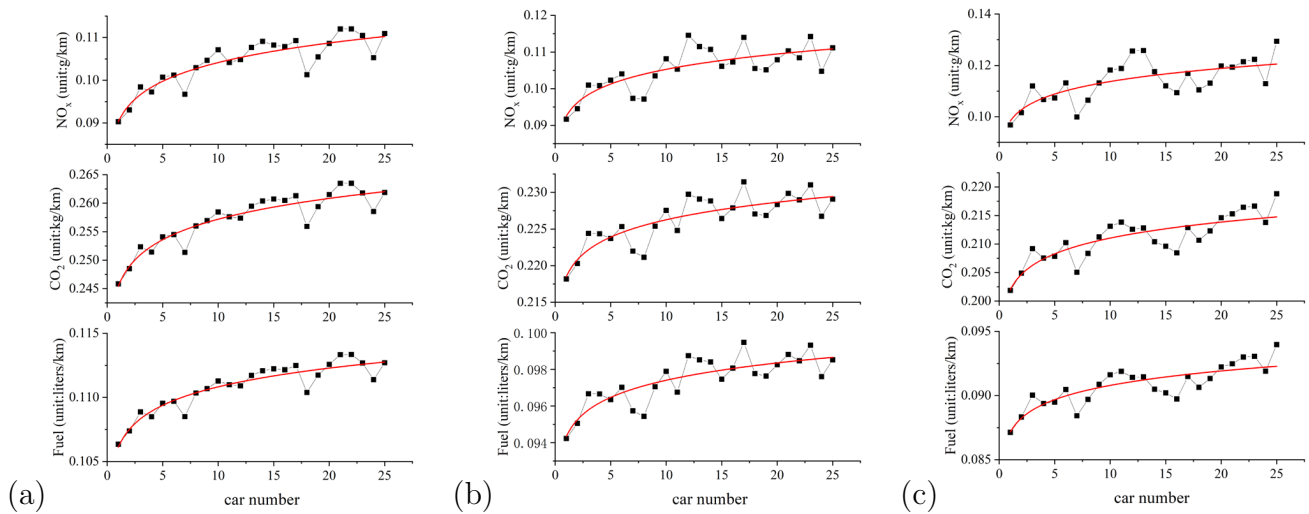

Figure 3: The experimental EFC of each car in the platoon calculated by the MEF model. The symbol solid black lines are the experiment results, and the red lines are the fitted lines. From (a) to (c), the velocity of the leading car moves with $v_{\text{leading}}$ =30, 40, and 50 km/h, respectively.

Figure 3 shows the EFC of each car along the platoon calculated by the MEF model, respectively. The results of other EFC models are shown in Appendix B. It can be seen that: (i) the fitting curve implies that all these indices exhibit a common feature of concave growth way along vehicles in the platoon; (ii) all the four EFC models exhibit consistent concave growth trends and show similar profiles of the EFC predictions. This indicates that the EFC also has a connection with traffic instability. The larger the car number is, the more EFCs are consumed. Additionally, in a long CF platoon, cars at the rear of the platoon will have similar EFCs because the concave growth gradually levels off.

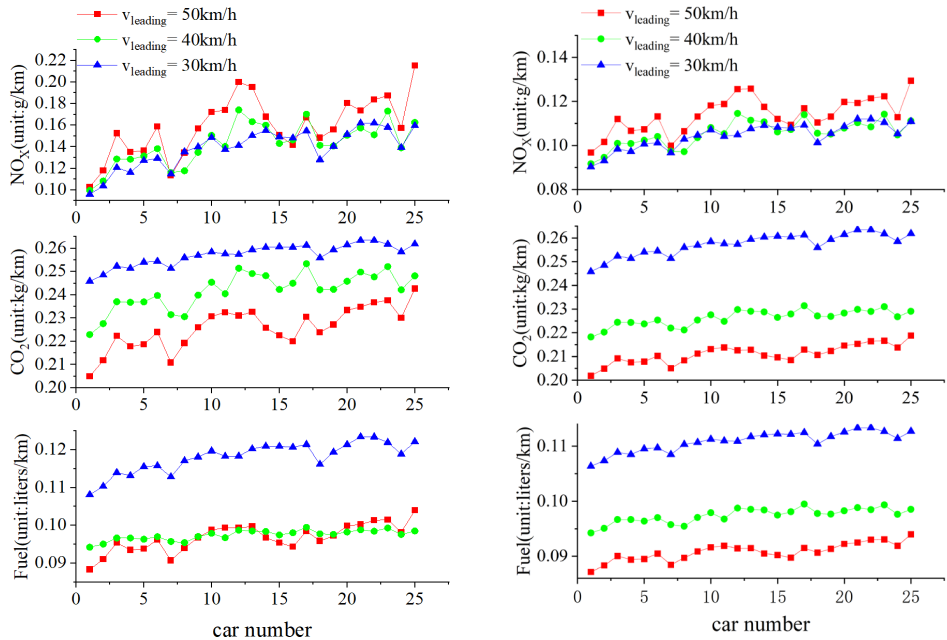

Figure 4: The experimental EFC of each car in the platoon with different $v_{\text{leading}}$ calculated by the VT-micro model and MEF model.

For further quantitative investigation on the impact of $v_{\text{leading}}$, Figure 4 plots all experimental EFCs of each car in the platoon. It can be seen that the difference between $NO_x$ is small under the three $v_{\text{leading}}$. In contrast, the emissions of $CO_2$ and fuel consumption significantly increases with the increase of $v_{\text{leading}}$, except that the fuel consumption calculated by the MEF model under $v_{\text{leading}}$= 50 and 40 km/h are similar. As stated in subsubsection 2.2.2, this may be because the MEF model incorporates comprehensive historical information into the composited acceleration, which makes the estimation of EFC under the high-speed condition (50 km/h) more accurate and robust than in the VT-micro model. To summarize, the universal characteristic of EFC, as demonstrated by our experiment with three $v_{\text{leading}}$ and four EFC models, is that they increase concavely in the platoon. There is no qualitative difference between the outcomes of the four EFC models.

## 3.2 Experiment II: 40-car platoon on a circular road

### 3.2.1 Experimental setup

On November 9, 2019, an experiment was conducted at the Research Institute of Highway test field under the Ministry of Transport in China. All cars involved in the experiment were equipped with high-precision GPS devices that recorded their locations and velocities every 0.1 seconds. The GPS devices had a measurement error within ±1 meter for location and ±1 km/h for velocity. Forty drivers and their cars were selected and divided into eight groups, labeled A through H. The number of groups and their order varied in different runs while maintaining a fixed order within each group.

Drivers were instructed to drive follow closely behind the vehicle ahead whenever it was safe to do so. At the beginning of each run, the cars were positioned bumper-to-bumper. The first car would then accelerate to 20 km/h and maintain that speed until all other cars started. Once all cars moved, the first car would catch up with the last car to complete the run. There are all six runs in this experiment. We choose one run for analysis.

### 3.2.2 Spatiotemporal properties in the circular platoon

The left panel in Figure 5 shows the spatiotemporal evolution of velocity and EFCs. Compared to Experiment I, the oscillations of Experiment II are nearly fully developed so that the profile of trajectories of different vehicles in the CF platoons are similar. The velocity and EFCs oscillate throughout the experiment. Many low-speed clusters emerge which indicates the stop-to-go traffic. Hence, experiment II can enrich the sceneries of EFC characteristics in oscillations, especially in stop-and-go waves. Moreover, the EFCs tend not to grow in the platoon, see Figure 6.

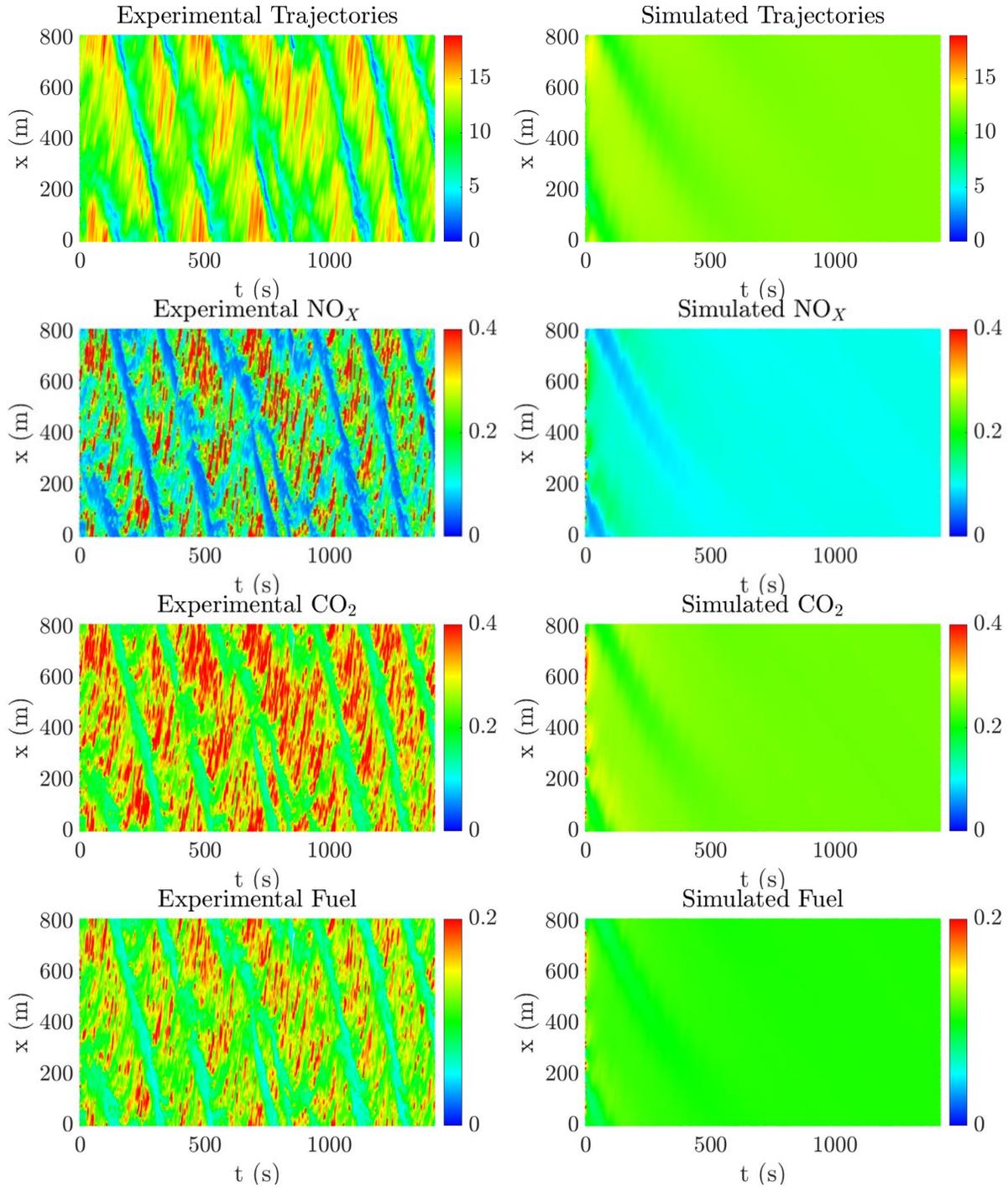

Figure 5: Evolution of the experimental (left) and simulated (right) spatiotemporal pattern of velocity and EFC of experiment II in platoon test. EFC is calculated by the MEF model with the calibrated IDM. From top to bottom, these panels represent velocity (km/h), $NO_x$ (mg), $CO_2$ (g), and Fuel (ml), respectively. In the simulation, only the initial velocity and the initial location of the vehicles are given.

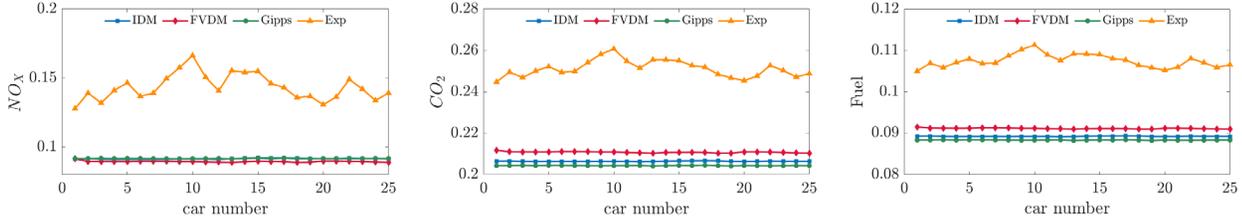

Figure 6: The EFC predictions along the platoons of experiment II in the platoon test. From left to right, these panels represent $NO_x$ (g/km), $CO_2$ (kg/km), and Fuel (L/km), respectively. The orange lines represent the experimental EFC of experiment II.

# 4 Model calibration and tests

To assess whether the current calibration framework and CF models can accurately reproduce the empirical properties of EFC, trajectory-based (local) calibration is conducted. Various simulations are conducted with calibrated parameters.

## 4.1 Calibration procedure

The key to an accurate estimation of EFC is to input the realistic velocity and acceleration into the EFC models. Hence, velocity is usually adopted as the measure of performance (see, e.g., Jie et al., 2013; Meng et al., 2021). However, according to Punzo and Montanino (2016), good calibration performance on velocity does not mean that the spacing is consistent with the observed data for single trajectory calibration because only the errors in spacing can keep the memory of model residuals dynamics while the errors on velocity cannot. Hence, in this paper, spacing is adopted for calibration. RMSPE is used as the goodness of fitness to identify the difference between experimental and simulated spacing. A genetic algorithm is applied to find the optimal solution by using the Matlab toolbox. The calibration boundaries are equal to the bounds of the vertical coordinate in Figure 7 and Figure 8. For each vehicle pair, we calibrate one set of parameters to take the heterogeneity of the driver into account. Hence, in terms of the object, the calibration procedure is based on the simulation from the local level.

## 4.2 Calibration results

The calibrated parameters of experiment I with different $v_{\text{leading}}$ have been shown in Figure 7. It demonstrates that (i) the parameters all show a large internal dispersion, and the parameters along the platoon have relatively large differences, which indicates the heterogeneity of driving behavior. (ii) Compared with IDM, Gipps has a smaller deceleration $b$ due to different CF mechanisms. (iii) the calibrated parameters are not significantly correlated to $v_{\text{leading}}$ except $v_0$. On the other hand, the parameters of experiment II have shown a relatively concentrated distribution. It might be due to the stop-to-go traffic making the trajectories similar to each other and the trajectories are longer than that in experiment I.

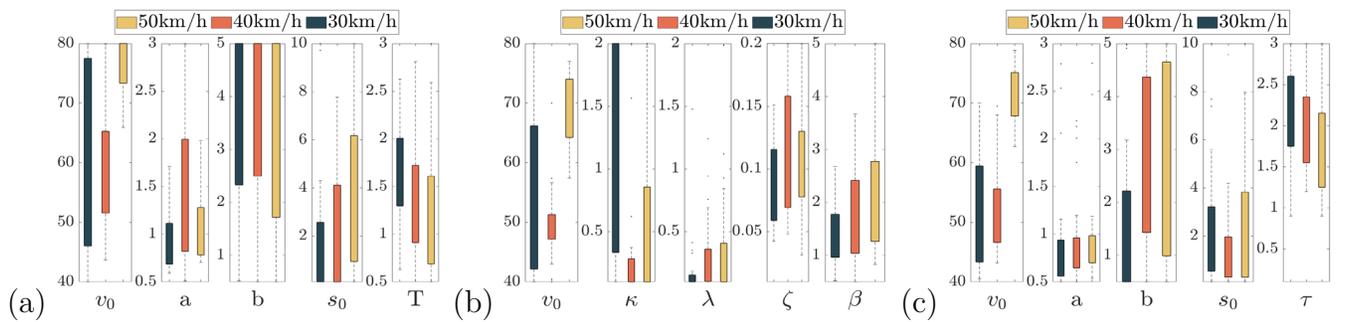

Figure 7: The distribution of calibrated parameters of (a)IDM, (b)FVDM, and (c)Gipps model in experiment I. The calibration ranges are the upper and lower bounds of the vertical coordinate of the boxplot.

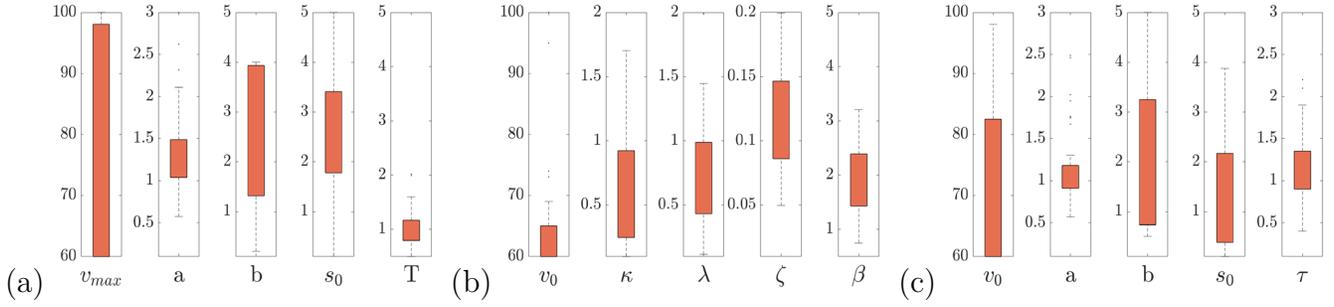

Figure 8: The distribution of calibrated parameters of (a)IDM, (b)FVDM, and (c)Gipps model in experiment II. The calibration ranges are the upper and lower bounds of the vertical coordinate of the boxplot.

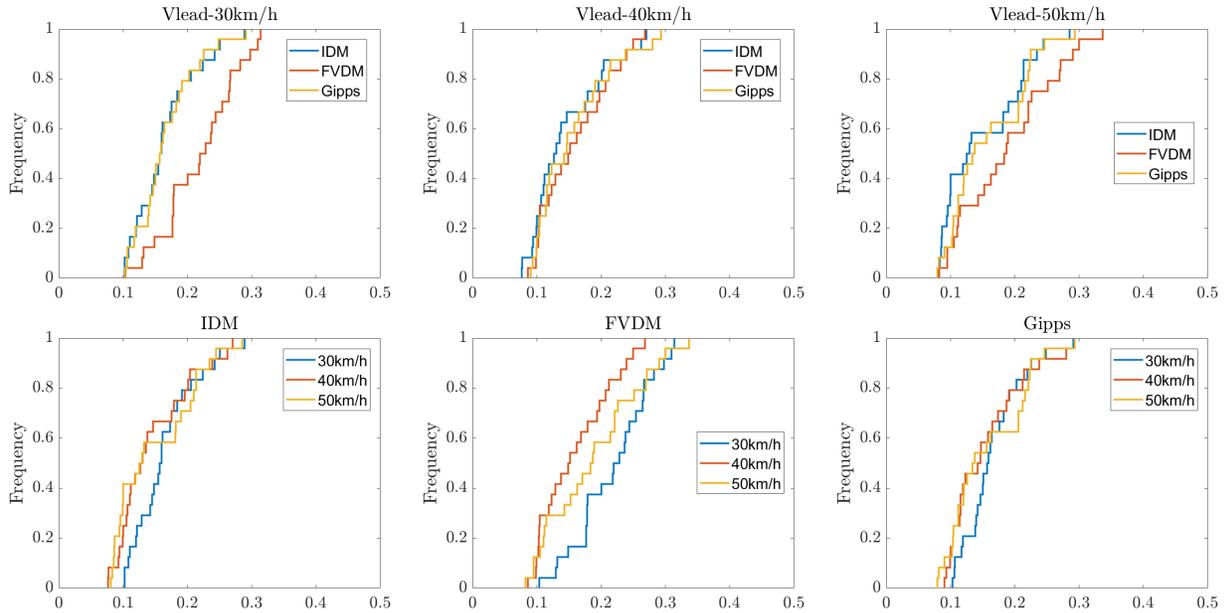

Figure 9: The cumulative frequency distribution of the calibration errors (RMSPE) of the spacing under different CF models and $\mathbf{v}_{leading}$ in experiment I.

Most of the calibration errors (RMSPE) are below 0.2, which indicates pretty good calibration performance. The calibration errors of experiment I on spacing are shown in Figure 9, which demonstrates (i) all three models perform pretty well from the perspective of calibration errors. (ii)

Comparing with the experiment under higher $v_{\text{leading}}$, the experiment of $v_{\text{leading}}$ =30 km/h has the relatively largest errors, especially for FVDM. (iii) Among the three CF models, the IDM and the Gipps model are better than FVDM. Furthermore, Figure 10 shows the results of experiment II, in which IDM performs best.

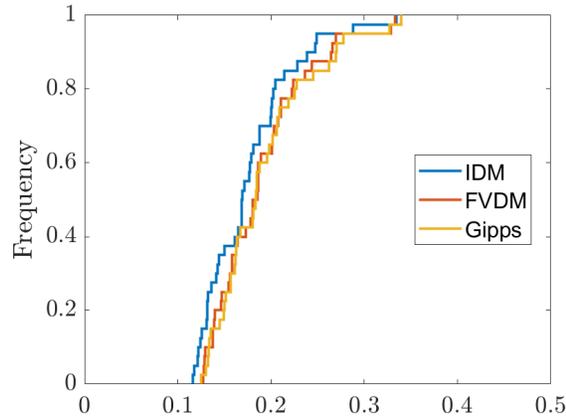

Figure 10: The cumulative frequency distribution of the calibration errors (RMSPE) of the spacing under different CF models in experiment II.

### 4.3   Test I: Local EFC prediction

To analyze the capability of calibrated models to predict EFC and reproduce their empirical properties, we first conduct a simulation test by identifying the local differences between the empirical and simulated EFC on the vehicle-pair level. For each pair of consecutive two vehicles, the real experimental trajectory of the leading vehicle is used as input to simulate the trajectory of the following vehicle.

Table 1: The EFC prediction error (RMSPE) in local and platoon tests in Experiment I.

| RMSPE(%) | $NO_x$ | | | $CO_2$ | | | Fuel | | |
|---|---|---|---|---|---|---|---|---|---|
| Local Test | IDM | FVDM | Gipps | IDM | FVDM | Gipps | IDM | FVDM | Gipps |
| 30km/h | 4.37 | 4.11 | 2.28 | 1.13 | 1.01 | 0.61 | 1.02 | 0.91 | 0.55 |
| 40km/h | 6.90 | 6.52 | 4.33 | 1.66 | 1.54 | 1.05 | 1.50 | 1.40 | 0.95 |
| 50km/h | 7.24 | 7.69 | 4.69 | 1.60 | 1.57 | 1.01 | 1.51 | 1.47 | 0.96 |
| Platoon Test | IDM | FVDM | Gipps | IDM | FVDM | Gipps | IDM | FVDM | Gipps |
| 30km/h | 10.60 | 9.81 | 10.15 | 3.81 | 3.59 | 3.53 | 3.49 | 3.21 | 3.22 |
| 40km/h | 13.57 | 12.89 | 10.44 | 3.99 | 3.69 | 3.17 | 3.61 | 3.36 | 2.89 |
| 50km/h | 12.52 | 13.59 | 9.62 | 3.59 | 3.60 | 3.17 | 3.31 | 3.27 | 2.86 |

Table 2: The EFC prediction error (RMSPE) in local and platoon tests in Experiment II.

| RMSPE(%) | $NO_x$ | | | $CO_2$ | | | Fuel | | |
|---|---|---|---|---|---|---|---|---|---|
| | IDM | FVDM | Gipps | IDM | FVDM | Gipps | IDM | FVDM | Gipps |
| Local Test | 6.92 | 6.00 | 6.67 | 1.33 | 1.24 | 1.37 | 1.24 | 1.15 | 1.29 |
| Platoon Test | 36.32 | 37.68 | 36.25 | 17.83 | 16.10 | 18.69 | 17.01 | 15.24 | 17.85 |

In Table 1, we find that compared with the calibration errors, the EFC prediction errors are largely smaller. This is probably because the calibration uses spacing as a measure while EFC uses velocity and acceleration. For all $v_{leading}$ and CF models, the $NO_x$ prediction is always significantly less accurate than that of the $CO_2$ and the fuel consumption. The prediction errors of $CO_2$ and Fuel are extremely low, ranging from 0.61-1.66%. It must be noted that even though the calibration errors of IDM on spacing are the smallest, its performance on EFC prediction error is reversed. The Gipps model has the smallest errors in the vehicle pair level.

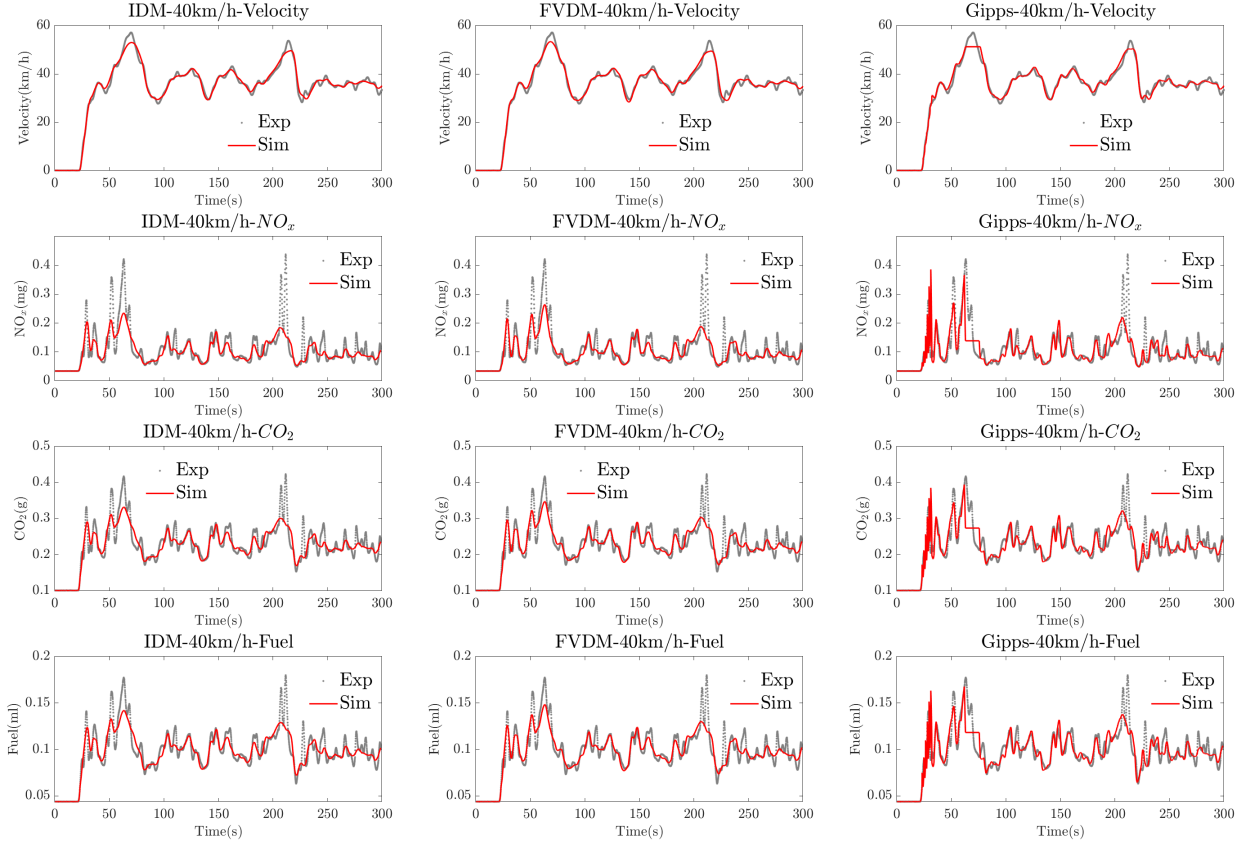

Figure 11: The simulated velocity and EFC of the $10^{\text{th}}$ car in experiment I under $v_{\text{leading}}$=40 km/h calculated by the MEF model with calibrated IDM, FVDM, and Gipps in the local test. The results of the other three EFC models are highly similar, hence not shown.

Figure 11 demonstrates the calibrated models of experiment I perform well in capturing CF and EFC dynamics at the vehicle-pair level. The velocity time series are accurately reproduced, but IDM and FVDM do not always successfully capture the peaks of the EFC time series due to inaccurate acceleration time series. In contrast, the Gipps model captures both velocity and EFC time series well, resulting in smaller prediction errors.

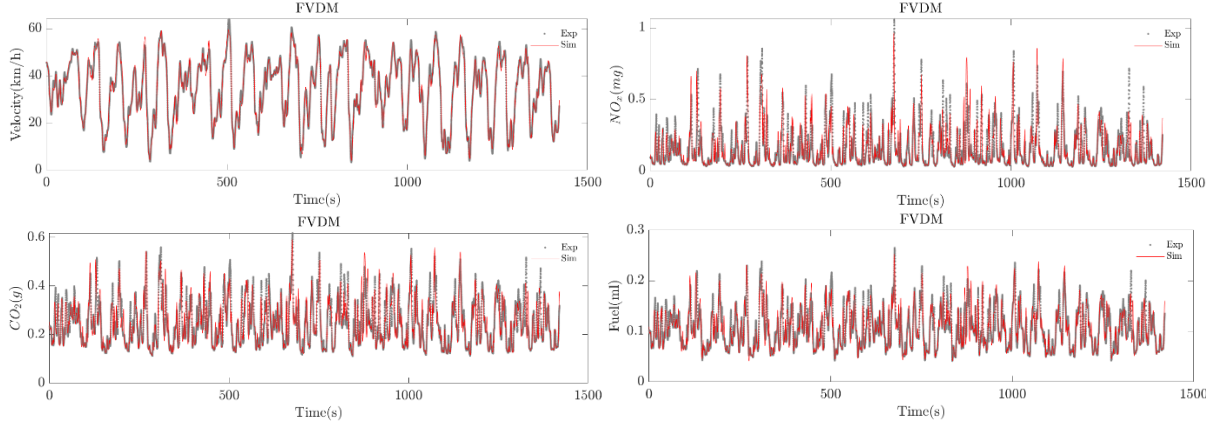

Figure 12: The simulated velocity and EFC of the 10$^{\text{th}}$ car in experiment II calculated by the MEF model with calibrated IDM, FVDM, and Gipps in the local test. The results of the other two CF models and three EFC models are highly similar, hence not shown.

In Experiment II, the sample simulated time series for velocity and EFCs are shown in Figure 12, which shows a high level of consistency with the experimental data. The local tests indicate that EFCs can be predicted with relatively small errors. However, the errors are still larger than those of Experiment I. Furthermore, it was consistently found that $NO_x$ had greater prediction errors than both $CO_2$ and Fuel. Among the three CF models, FVDM performs best in reproducing both velocity and EFC.

To summarize, no matter how the oscillation develops, the calibrations can reproduce the real EFC with small errors.

### 4.4 Test II: Platoon EFC prediction

In the urban transportation simulations, it is necessary to investigate the EFC prediction at the platoon level. All the vehicles are simulated at the same time in the simulation procedure.

For the experiment I, Figure 13 illustrates the spatiotemporal evolution of EFC simulated by the three calibrated CF models, which are qualitatively different from the experimental one. It shows that the EFC prediction has no oscillation property like Figure 2 and is nearly smooth (see Figure 14). Table 1 also shows the EFC prediction results in the platoon test, indicating that none of the CF models can accurately estimate the EFC of the experimental trajectory. Particularly, the largest relative error on $NO_x$ is more than 13%. The relative error of platoon test on $CO_2$ and fuel are nearly triple that of local tests. Moreover, although the EFC grows in the platoon in Figure 15, the predictions are not accurate. Hence, the current framework may not be efficient for EFC predictions at the platoon level.

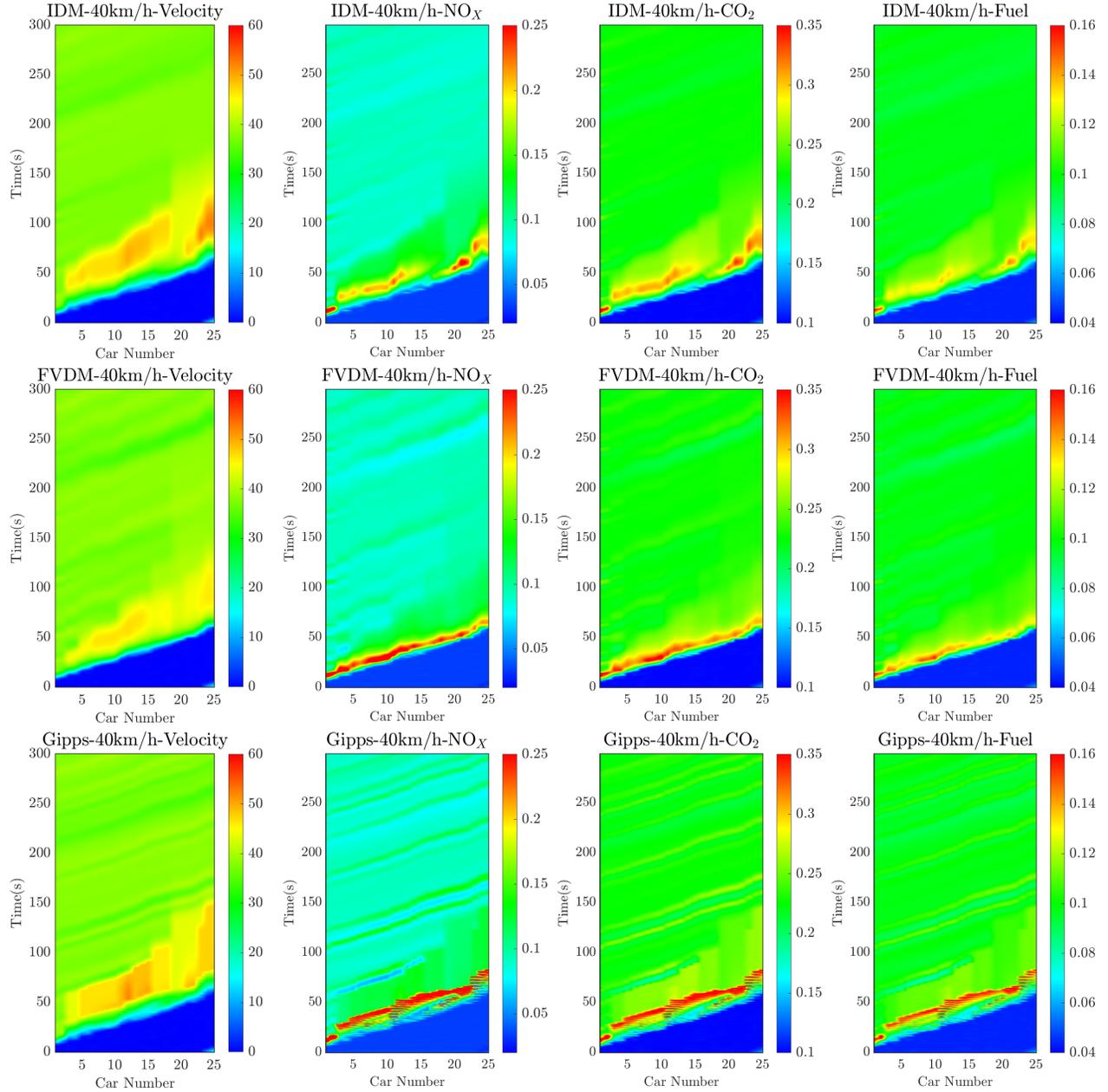

Figure 13: Evolution of the spatiotemporal pattern of velocity and EFC in experiment I in platoon test when $v_{\text{leading}}$=40km/h. EFC is calculated by the MEF model (the results of the other three EFC models are highly similar, hence not shown). From left to right, these panels represent velocity (km/h), $NO_x$ (mg), $CO_2$ (g), and Fuel (ml), respectively. In the simulation, the trajectory of the leading car is given. For the following cars, only the initial velocity and the initial location are given.

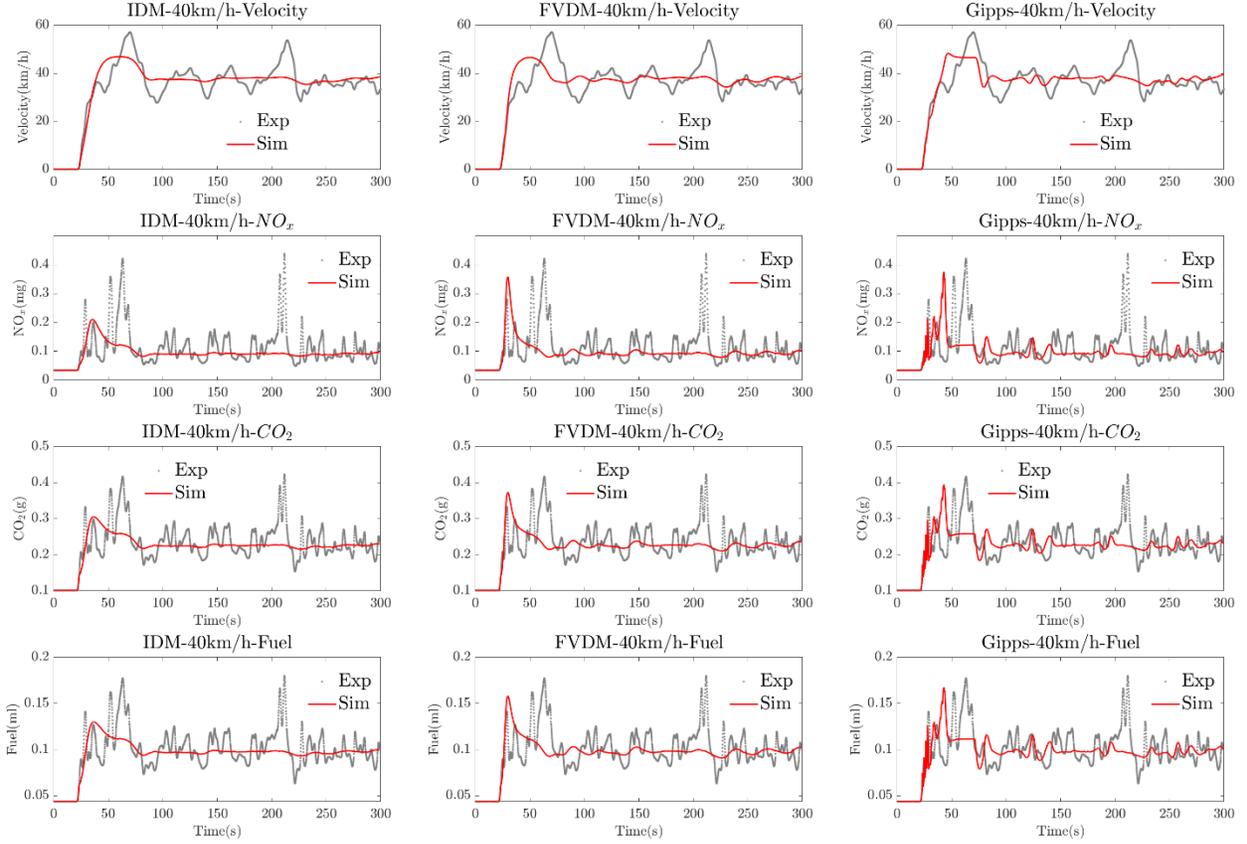

Figure 14: The simulated velocity and EFC of the $10^{\text{th}}$ car in experiment I under $v_{\text{leading}}$=40 km/h calculated by the MEF model with calibrated IDM, FVDM, and Gipps in the platoon test. The results of the other three EFC models are highly similar, hence not shown.

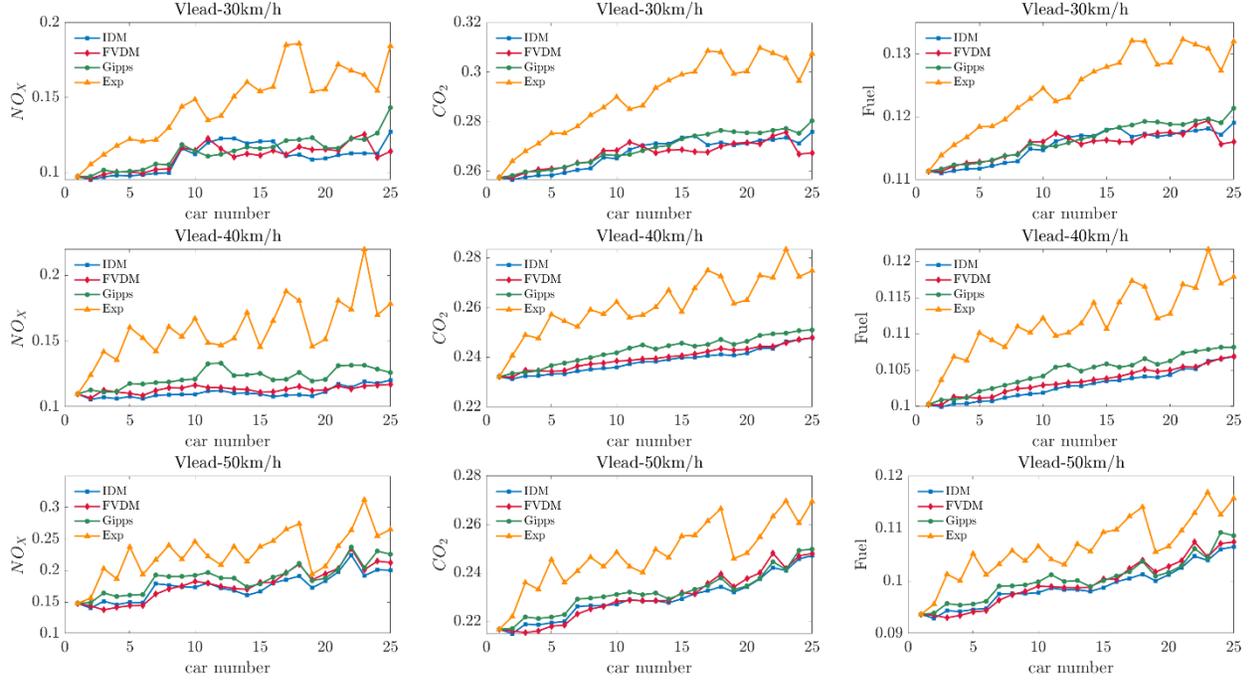

Figure 15: The EFC predictions along the platoons in the platoon test calculated by the MEF model (the results of the other three EFC models are highly similar, hence not shown). From left to right, these panels represent $NO_x$ (g/km), $CO_2$ (kg/km), and Fuel (L/km), respectively.

These phenomena are more pronounced in the results of experiment 2, see the right panels in Figure 5. The EFC prediction errors rise rapidly. For instance, the error of IDM on $NO_X$ in the platoon test is 36.32% while it is 6.92% in the local test, see Table 2. This indicates that there may exist quantitative differences between simulated and real trajectories at the platoon level. In Figure 6, the simulated spatiotemporal diagrams show no oscillations not only in velocity but also in all EFCs. It can be clearer in Figure 6, the predicted EFC is much smaller than the real one and is nearly constant along the platoon. The main reason is that the traffic flow tends to be homogeneous where the oscillations are not captured by the calibrated CF models.

# 5    Conclusion

Accurate EFC assessment is essential for sustainable transportation systems and forms the basis for eco-driving control. For decades, CF models have been coupled with EFC models to generate velocity and acceleration series, which are then used as input for the EFC models to estimate EFC. However, the EFC property of car platoons has been neglected for a long time. To address this research gap, the experimental data of the 25-car platoon on a straight road and the 40-car platoon on a circular road are analyzed using four EFC models.

These two experiments reflect the different stages of oscillations development. In experiment I, the results showed a concave growth pattern of EFC along the platoon for all the EFC models, suggesting that this is a universal phenomenon. It means the EFC rises rapidly in the CF platoon and then tends to be flattening. On the other hand, for the fully developed oscillations in experiment II, there is no significant difference among the EFCs of different vehicles in the platoon.

We calibrate three CF models to evaluate the current framework of EFC prediction. On this basis, two tests are conducted from the vehicle-pair level and platoon level, respectively. For the results of experiments I and II in local EFC prediction, all vehicle pairs successfully capture the CF dynamics and predict the EFC with nearly 1% errors on $CO_2$ and fuel while $NO_x$ has relatively larger errors.

However, the vehicle-pair level is not able to represent the property of CF dynamics and EFC in the platoon system. We further conduct the platoon EFC prediction. Qualitative differences are observed at the platoon level. The EFC is nearly constant in the platoon and does not grow. The spatiotemporal evolution shows no obvious velocity and EFC oscillation. The largest prediction error is more than 37%. Hence, the current framework may not be able to predict the EFC in the platoon with velocity oscillation.

In the future, a systematic framework of EFC predictions should be established from the perspective of the selection of CF models, EFC models, and calibration procedures, see Figure 1. Additionally, a more detailed microscopic analysis of velocity and acceleration series should be conducted to provide direction for developing CF models, allowing them to accurately replicate the spatiotemporal patterns of CF dynamics and EFC in the platoon.

# 6 Acknowledgments

This work is supported by the National Natural Science Foundation of China (Grant No. 72222021, 72288101, 71931002, 72010107004).

# Appendix A

Table A1: The MOE fuel consumption Coefficients (the units of speed, acceleration, and fuel consumption are km/h, km/h/s, and liters/s respectively)

| $K_{ij}$ | $a_n(t) > 0$ | | | | $a_n(t) \leq 0$ | | | |
|---|---|---|---|---|---|---|---|---|
| | j = 0 | j = 1 | j = 2 | j = 3 | j = 0 | j = 1 | j = 2 | j = 3 |
| i = 0 | -7.735 | 0.2295 | -5.61E-03 | 9.77E-05 | -7.735 | -0.01799 | -4.27E-03 | 1.88E-04 |
| i = 1 | 0.02799 | 0.0068 | -7.72E-04 | 8.38E-06 | 0.02804 | 7.72E-03 | 8.38E-04 | 3.39E-05 |
| i = 2 | -2.23E-04 | -4.40E-05 | 7.90E-07 | 8.17E-07 | -2.20E-04 | -5.22E-05 | -7.44E-06 | 2.77E-07 |
| i = 3 | 1.09E-06 | 4.80E-08 | 3.27E-08 | -7.79E-09 | 1.08E-06 | 2.47E-07 | 4.87E-08 | 3.79E-10 |

Table A2: The MOE $CO_2$ emission Coefficients (the units of speed, acceleration, and emission are km/h, km/h/s, and mg/s respectively).

| $K_{ij}$ | $a_n(t) > 0$ | | | | $a_n(t) \leq 0$ | | | |
|---|---|---|---|---|---|---|---|---|
| | j = 0 | j = 1 | j = 2 | j = 3 | j = 0 | j = 1 | j = 2 | j = 3 |
| i = 0 | 6.916 | 0.217 | 2.35E-04 | -3.64E-04 | 6.915 | -0.032 | -9.17E-03 | -2.89E-04 |
| i = 1 | 0.02754 | 9.68E-03 | -1.75E-03 | 8.35E-05 | 0.0284 | 8.53E-03 | 1.15E-03 | -3.06E-06 |
| i = 2 | -2.07E-04 | -1.01E-04 | 1.97E-05 | -1.02E-06 | -2.27E-04 | -6.59E-05 | -1.29E-05 | -2.68E-07 |
| i = 3 | 9.80E-07 | 3.66E-07 | -1.08E-07 | 8.50E-09 | 1.11E-06 | 3.20E-07 | 7.56E-08 | 2.95E-09 |

Table A3: The MOE $NO_x$ emission Coefficients (the units of speed, acceleration, and emission are km/h, km/h/s, and mg/s respectively).

| $K_{ij}$ | $a_n(t) > 0$ | | | | $a_n(t) \leq 0$ | | | |
|---|---|---|---|---|---|---|---|---|
| | j = 0 | j = 1 | j = 2 | j = 3 | j = 0 | j = 1 | j = 2 | j = 3 |
| i = 0 | -1.08 | 0.2369 | 1.47E-03 | -7.82E-05 | -1.08 | 0.2085 | 2.19E-02 | 8.82E-04 |
| i = 1 | 1.79E-02 | 4.05E-02 | -3.75E-03 | 1.05E-04 | 2.11E-02 | 1.07E-02 | 6.55E-03 | 6.27E-04 |
| i = 2 | 2.41E-04 | -4.08E-04 | -1.28E-05 | 1.52E-06 | 1.63E-04 | -3.23E-05 | -9.43E-05 | -1.01E-05 |
| i = 3 | -1.06E-06 | 9.42E-07 | 1.86E-07 | 4.42E-09 | -5.83E-07 | 1.83E-07 | 4.47E-07 | 4.57E-08 |

# Appendix B

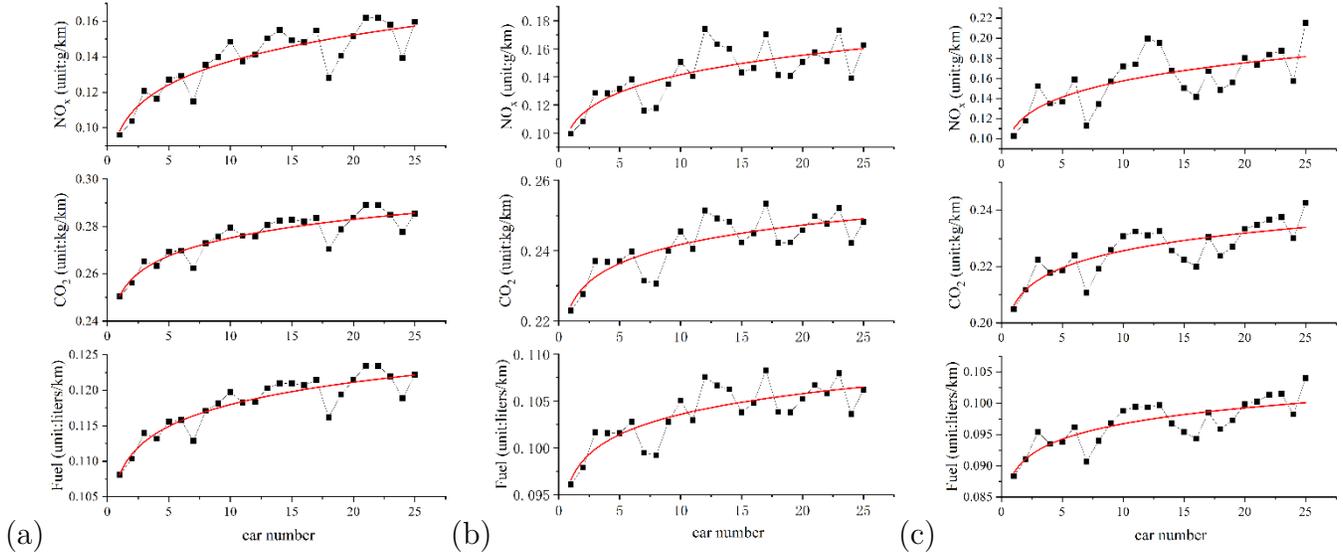

Figure B1: The experimental EFC of each car in the platoon using the VT-Micro model. The symbol solid black lines are the experiment results and the red lines are the fitted lines. From (a) to (c), the velocity of the leading car moves with $\mathbf{v}_{leading}$ =30, 40, and 50 km/h respectively.

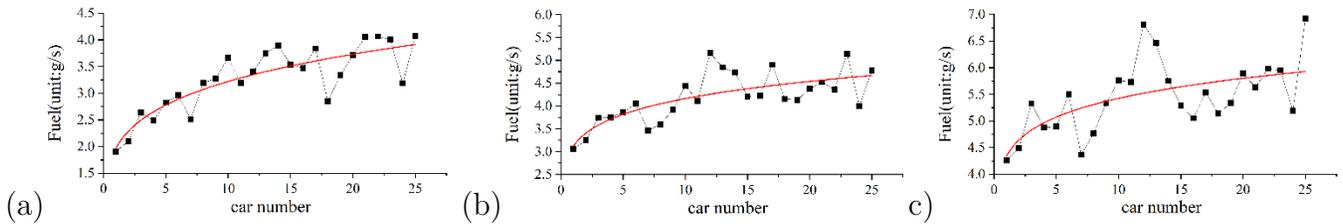

Figure B2: The experimental EFC of each car in the platoon using the ARRB model. The symbol solid black lines are the experiment results and the red lines are the fitted lines. From (a) to (c), the velocity of the leading car moves with $\mathbf{v}_{leading}$ =30, 40, and 50 km/h respectively.

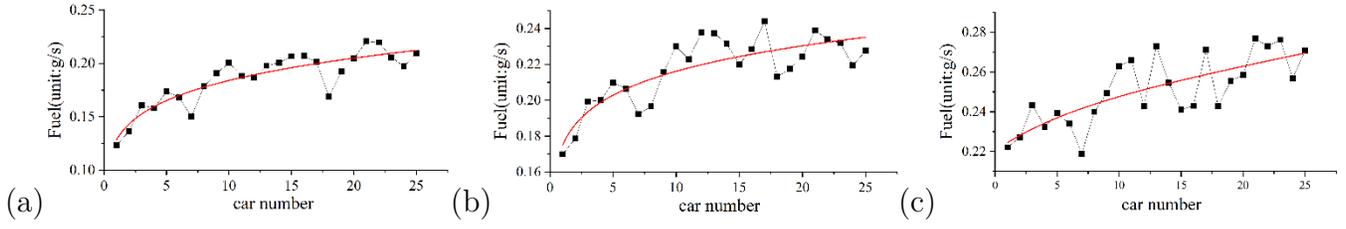

Figure B3: The experimental EFC of each car in the platoon using the VSP model. The symbol solid black lines are the experiment results and the red lines are the fitted lines. From (a) to (c), the velocity of the leading car moves with $\mathbf{v}_{leading}$ =30, 40, and 50 km/h respectively.